\documentclass[%
 reprint,
 amsmath,amssymb,
 aps,
 prl,
]{revtex4-2}
 
\usepackage{graphicx}
\usepackage{dcolumn}
\usepackage{bm}
\usepackage{float}
\usepackage{xcolor}

\begin{document}
 
\preprint{APS/123-QED}
 
\title{Cold kinetic theory is not cold fluid theory:\\ A collective mode in self-gravitating spheres}
 
\author{Mikael Tacu}
 \email{mikael.tacu@gmail.com}
\affiliation{CEA, DAM, DIF, F-91297 Arpajon, France}
\affiliation{Universit\'{e} Paris-Saclay, CEA, Laboratoire
Mati\`{e}re en Conditions Extr\^{e}mes, F-91680
Bruy\`{e}res-le-Ch\^{a}tel, France}
\date{\today}

\begin{abstract}
Cold collisionless matter is usually described by pressureless fluid equations; this fails in three-dimensional spherical geometry. For the cold spherical case, kinetic and fluid descriptions share the same time-domain dynamics, but if we look at the spectrum, the kinetic system admits one solution the fluid does not: a discrete mode at twice the central orbital frequency, $\omega_0 = 2\Omega(0)$. It is specific to 3D: razor-thin disks reduce to the fluid description. Nonlinear simulations confirm it survives at finite dispersion. Around Sgr~A$^*$ it lies at $\kappa(r_{\rm in})$, with $r_{\rm in}$ the cusp's inner edge; if excited, stars at all radii would oscillate at this one frequency.
\end{abstract}

\maketitle

It is standard practice to describe cold collisionless matter as a pressureless fluid. For a zero temperature distribution, the velocity moments of the Vlasov equation close exactly onto the pressureless Euler equations~\cite{binney2008,peebles1980}, and the closure remains exact for cold orbit-supported equilibria, each mass shell conserving its angular momentum. The fluid treatment of cold
dark-matter dynamics~\cite{peebles1980}, cold stellar
disks~\cite{LinShu1964,Toomre1964}, and cold gravitational collapses~\cite{fillmore1984} relies on this equivalence, which is
regarded as exact.

That the two descriptions might differ is not obvious: in one spatial
dimension, the cold kinetic and fluid equations are indeed equivalent.
The same is true for razor-thin disks, as we show below.  The failure is
specific to three-dimensional spherical geometry, where the
velocity-space measure carries a factor of angular momentum that has
no counterpart in lower-dimensional systems.  This factor prevents the
Poisson integral from reducing to a local relation between the
perturbed potential and the displacement, so that the kinetic system
admits a discrete collective mode that the fluid equations do not.
 
In what follows we derive the mode analytically, identify the physical
mechanism that selects its frequency, prove its absence in disks,
confirm it with nonlinear Vlasov--Poisson simulations, and apply the
result to the stellar cusp around Sgr~A$^*$.
 
\textit{Equilibrium.}---Consider a spherically symmetric potential
$\psi_0(r)$ and the cold distribution
\begin{equation}
f_0 = g(r)\delta(v_r)\delta(v_\perp - \sqrt{r\psi_0'}),
\label{eq:f0}
\end{equation}
which places all stars on circular orbits with transverse velocity
$v_\perp = \sqrt{v_\theta^2+v_\varphi^2} = \sqrt{r\psi_0'}$.
This is a stationary solution of the spherical Vlasov equation
\begin{equation}
\frac{\partial f}{\partial t} + v_r\frac{\partial f}{\partial r}
+ \left(\frac{v_\perp^2}{r} - \frac{\partial \psi}{\partial r}\right)
\frac{\partial f}{\partial v_r}
- \frac{v_r v_\perp}{r}
\frac{\partial f}{\partial v_\perp} = 0.
\label{eq:vlasov}
\end{equation}
Ng and Bhattacharjee~\cite{Ng2005,Ng2006} constructed 3D equilibria using $\delta(v_\perp^2/r-\psi_0')$, which is ambiguous since by definition $v_\perp \ge 0$ .  Our formulation~\cite{Tacu2025} with $\delta(v_\perp-\sqrt{r\psi_0'})$ removes that ambiguity and makes the perturbative analysis tractable. 

Introducing the angular momentum $\ell = rv_\perp$ and the profile $\xi(r) = r\sqrt{r\psi_0'}$, the Poisson equation reads
\begin{equation}
\partial_r
\left(r^2\partial_r\psi\right)
= 8\pi^2 G \int_{\mathbb{R}}dv_r
\int_0^{\infty}\ell d\ell f.
\label{eq:poisson_eq}
\end{equation}
Evaluating on the equilibrium~\eqref{eq:f0}, it reduces to
\begin{equation}
r\psi_0'' + 2\psi_0' = 8\pi^2 G\xi g. 
\label{eq:equil}
\end{equation}
\textit{Linearized kinetic system.}---We restrict to spherically symmetric perturbations and pass to the variables
$(r, v_r, \ell)$, since the angular momentum is an
exact invariant in spherical symmetry.  It parametrizes a family of independent 1D-1V problems in $(r, v_r)$ coupled only through the
Poisson equation.  Setting $F(r,v_r,\ell,t) = f(r,v_r,rv_\perp,t)$,
the linearized Vlasov equation reads
\begin{equation}
\frac{\partial F_1}{\partial t} + v_r\frac{\partial F_1}{\partial r}
+ \frac{\ell^2 - \xi^2}{r^3}
\frac{\partial F_1}{\partial v_r}
= rg\psi_1'\delta'(v_r)\delta(\ell{-}\xi).
\label{eq:lin_vlasov}
\end{equation}
This can be solved using the following ansatz with three fields
$a_i(r,t)$,
\begin{equation}
F_1 = a_1\delta(v_r)\delta'(\ell{-}\xi)
+ a_2\delta'(v_r)\delta(\ell{-}\xi)
+ a_3\delta(v_r)\delta(\ell{-}\xi).
\label{eq:ansatz}
\end{equation}
Substituting it into Eq.~\eqref{eq:lin_vlasov} and using the  identities $(\ell{-}\xi)\delta'(\ell{-}\xi) = -\delta(\ell{-}\xi)$
and $v_r\,\delta'(v_r) = -\delta(v_r)$, one collects the
coefficients of each independent distribution and obtains
\begin{align}
\partial_t a_1 &= -\xi'a_2,
\label{eq:sys1}\\
\partial_t a_2 - \frac{2\xi}{r^3}a_1 &= rg\psi_1',
\label{eq:sys2}\\
\partial_t a_3 &= \partial_r a_2.
\label{eq:sys3}
\end{align}
%
\textit{Analytical solution.}---The
system~\eqref{eq:sys1}--\eqref{eq:sys3} can be solved in closed
form.  Define $\alpha(r,t)$ by $a_2 = \partial_t\alpha$.
Equations~\eqref{eq:sys1} and~\eqref{eq:sys3} integrate to
\begin{equation}
a_1 = -\xi'\alpha,\qquad a_3 = \partial_r\alpha.
\label{eq:alpha_field}
\end{equation}
The perturbed Poisson equation becomes,
\begin{equation}
\partial_r(r^2\partial_r\psi_1) = 8\pi^2 G (\xi a_3 - a_1)
= 8\pi^2G\partial_r(\xi \alpha).
\label{eq:poisson_pert}
\end{equation}
Integrating from $0$ to $r$:
\begin{equation}
r^2\partial_r \psi_1 = 8\pi^2 G\xi\alpha + K.
\label{eq:poisson_K}
\end{equation}
Regularity at the origin forces $K = 0$.  Substituting
$\psi_1' = 8\pi^2 G\xi\alpha/r^2$ into~\eqref{eq:sys2} and
using the equilibrium relation~\eqref{eq:equil} to replace
$8\pi^2 G\xi g$:
\begin{equation}
\ddot\alpha + \Omega^2(r)\alpha = 0,\qquad
\Omega = \sqrt{\psi_0'/r}.
\label{eq:alpha_osc}
\end{equation}
The kinetic system reduces to uncoupled harmonic oscillators.
The solution is
\begin{equation}
\alpha(r,t) = A(r)\sin\bigl[\Omega(r)t + \phi(r)\bigr],
\label{eq:alpha_sol}
\end{equation}
with $A(r)$ and $\phi(r)$ set by initial conditions.  Each shell
oscillates independently at the local orbital frequency $\Omega(r)$.
The oscillations dephase across shells, and the resulting signal
decays by phase mixing.
 
\textit{Fluid equations.}---The exact fluid counterpart of~\eqref{eq:f0} is a nest of cold shells, each conserving its angular momentum and thereby carrying the equilibrium tangential stress. The radial displacement $\delta r$ of a shell obeys
\begin{equation}
\ddot{\delta r} + \kappa^2\delta r = -\partial_r\psi_1,
\label{eq:euler}
\end{equation}
with $\kappa^2 = 3\psi_0'/r + \psi_0''$ the epicyclic frequency
squared, the first term arising from angular-momentum conservation. Mass conservation gives for the perturbed density
$\rho_1 = -r^{-2}\partial_r(r^2\rho_0\,\delta r)$, which is a total
derivative.  Inserting into the Poisson equation and integrating:
\begin{equation}
r^2\psi_1' = -4\pi Gr^2\rho_0\delta r + C.
\label{eq:poisson_fluid}
\end{equation}
Regularity at the origin forces $C = 0$.  Substituting back
into~\eqref{eq:euler}:
\begin{equation}
\ddot{\delta r} + \Omega^2(r)\,\delta r = 0.
\label{eq:fluid}
\end{equation}
This is identical to the kinetic result~\eqref{eq:alpha_osc}.
Each shell oscillates independently at $\Omega(r)$; the fluid
spectrum is purely continuous.  In the time domain, the two systems
are indistinguishable.
 
\textit{Normal-mode analysis.}---The difference appears in the
spectral domain.  For eigenmodes $\psi_1' = z(r)e^{i\omega t}$,
Eqs.~\eqref{eq:sys1}--\eqref{eq:sys3} give
$a_2 = -i\omega rg\,z/(\omega^2 - \kappa^2)$, and the combination
entering the density satisfies
\begin{equation}
\xi a_3 - a_1 = -\frac{i}{\omega}
\partial_r(\xi a_2).
\label{eq:total_deriv}
\end{equation}
Substituting into~\eqref{eq:poisson_pert} yields
$\partial_r(r^2 z)
= -(8\pi^2 G i/\omega)\partial_r(\xi a_2)$.
Integrating from $0$ to $r$ and substituting $a_2$:
\begin{equation}
r^2 z = - \frac{8\pi^2 G r g \xi}
{\omega^2 - \kappa^2} z + K,
\label{eq:intermediate}
\end{equation}
where $K$ is an integration constant.  Using the equilibrium
relation~\eqref{eq:equil} to replace $8\pi^2 G \xi g$
by $2\psi_0' + r\psi_0''$, this simplifies to
\begin{equation}
\alpha(r) z(r) = K,
\label{eq:conserved}
\end{equation}
with
\begin{equation}
\alpha(r) = r^2 + \frac{r(2\psi_0'+ r\psi_0'')}
{\omega^2 - \kappa^2(r)}.
\label{eq:alpha_def}
\end{equation}
The fluid equations yield the same
relation~\eqref{eq:conserved} with the same $\alpha$, but with
$C$ in place of $K$.  In the time-domain reduction above, regularity
forced $K = 0$ at all times. The
question is whether a single frequency $\omega$ admits $K \neq 0$
with $z$ regular at the origin.
 
Using $r(2\psi_0' + r\psi_0'') = 4\pi G\rho_0 r^2$ and
$\kappa^2 - 4\pi G\rho_0 = \Omega^2$,
Eq.~\eqref{eq:alpha_def} takes the form
\begin{equation}
\alpha(r) = r^2 \frac{\omega^2 - \Omega^2(r)}
{\omega^2 - \kappa^2(r)} .
\label{eq:alpha_ratio}
\end{equation}
In the fluid, regularity forces $C = 0$ at every $\omega$: the
spectrum is the continuum $\{\Omega(r)\}$.  In the
kinetic system, $K \neq 0$ is admissible if and only if
$z = K/\alpha$ is regular at the origin, which requires
$\alpha(0) \neq 0$.
 
\textit{Discrete frequency.}---Near $r = 0$ the potential is
harmonic: $\psi_0' = \Omega_0^2r + O(r^3)$ with
$\Omega_0^2 \equiv \psi_0''(0) = \Omega^2(0)$ and $\kappa^2(0) = 4\Omega_0^2$.  For generic
$\omega$, $\alpha \propto r^2$ and $z = K/\alpha \sim 1/r^2$.
The perturbed enclosed mass $r^2\psi_1' \propto r^2 z$ stays
finite at the origin, violating Gauss's theorem which requires
$r^2\psi_1' \to 0$.
 
\begin{figure}[t]
\centering
\includegraphics[width=\columnwidth]{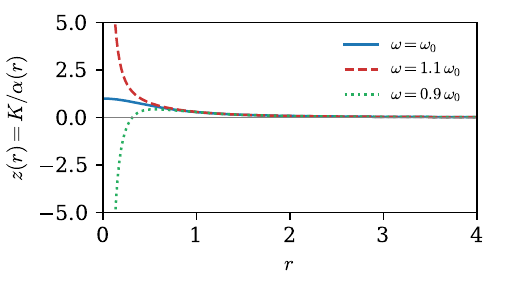}
\caption{%
Mode profile $z(r) = K/\alpha(r)$ for a Plummer potential
($r$ in units of~$a$).  At the
discrete frequency $\omega = \omega_0 = \kappa(0)$ (solid blue),
$z$ is finite at the origin and decays as $1/r^2$ at large $r$.
At $\omega \neq \omega_0$ (dashed red, dotted green), $z$ diverges
as $1/r^2$ at the origin, so no regular mode exists at any other
frequency.}
\label{fig:mode}
\end{figure}

Regularity demands that $\omega^2 - \kappa^2(r)$ vanish at the
same rate as $r^2$.  Since
$\kappa^2 = 4\Omega_0^2 - |\psi_0^{(4)}(0)|\,r^2 + O(r^4)$, with $\psi_0^{(4)}(0)<0$ for a centrally peaked density, this selects
\begin{equation}
\boxed{\omega_0 = 2\sqrt{\psi_0''(0)} = \kappa(0),}
\label{eq:omega0}
\end{equation}
where $\alpha \to -3\Omega_0^2/\psi_0^{(4)}(0) \neq 0$ and $z$
is finite (Fig.~\ref{fig:mode}).  The fluid displacement
$\delta r = z/(\omega_0^2 - \kappa^2) \sim 1/r^2$ diverges, so
the fluid rejects the mode, while the kinetic system, whose perturbed force and enclosed mass remain regular at the origin, accepts it. For a Plummer model ($\psi_0 = -GM/\sqrt{r^2+a^2}$),
$\omega_0 = 2\sqrt{GM/a^3} = 2\,\Omega(0)$.

\textit{Physical mechanism.}---%
The continuum and the discrete mode are the singular ($K=0$) and regular ($K\neq0$)
solutions of the spectral relation~\eqref{eq:conserved}. For $K=0$ the only regular solution is $z\equiv0$: the continuum is carried by singular van~Kampen modes\cite{Mathur1990}: $z_\omega(r)\propto\delta(r-r_*)$, $\Omega(r_*)=\omega$, each localized on
one shell. There $\alpha(r_*)=0$, so $\alpha z=0$ is satisfied with $K=0$. The cold shell has a single orbital frequency, so the singularity is a pure $\delta$ with no principal-value part. Indexed by $\omega\in(0,\Omega_0]$ they form the continuum $\{\Omega(r)\}$, and their superposition is the time-domain solution, which decays by phase-mixing (the gravitational analogue of Landau phase-mixing). For $K\neq0$, $z=K/\alpha$ is a single regular profile oscillating at one frequency on every shell: the Poisson integral ties the shells together. Two regularity conditions then select the admissible $\omega$. First, $z=K/\alpha$ is singular wherever $\alpha=0$, i.e.\ wherever $\omega=\Omega(r)$, so a regular mode requires $\omega$ above the
continuum: $\omega>\Omega_0\equiv\max\Omega(r)=\Omega(0)$. Second, at the origin $\alpha\propto r^2$ for generic $\omega$, giving $z\sim 1/r^2$ and a finite perturbed enclosed mass $r^2\psi_1'$ in violation of Gauss's theorem; $\alpha(0)\neq0$ requires $\omega^2-\kappa^2(r)$ to vanish as $r^2$, which selects $\omega_0=\kappa(0)=2\Omega_0$
(Eq.~\eqref{eq:omega0}). This lies at twice the top of the continuum: $\Omega(r)<\omega_0$ for every $r$, the mode meets no resonance, and it is undamped in the cold limit. It is worth mentioning that $z(r)  \propto z_\infty/r^2$ when $r\to \infty$, which is the physical behavior.

\textit{Absence in disks.}---In a razor-thin disk the velocity-space
measure is $dv_r\,d\ell/r$ with no  $\ell$ prefactor.  The 2D Vlasov
equation in $(r,\theta,v_r,\ell)$,
\begin{equation}
\partial_t F + v_r\partial_r F
+ \tfrac{\ell}{r^2}\partial_\theta F
+ (\tfrac{\ell^2}{r^3} - \tfrac{\partial \psi}{\partial r})\partial_{v_r}F
- \partial_\theta\psi\partial_\ell F = 0,
\label{eq:vlasov2d}
\end{equation}
admits the equilibrium $F_0 = rg\,\delta(v_r)\delta(\ell{-}\xi)$
with $4\pi G g
= \psi_0'' + \psi_0'/r$ by Poisson equation.  For axisymmetric perturbations ($\partial_\theta = 0$), the linearized Vlasov equation can be written as, 
\begin{equation}
\begin{gathered}
\partial_t F_1 + v_r\partial_r F_1 + \tfrac{\ell}{r^2}\partial_\theta F_1 + \tfrac{(\ell-\xi)(\ell+\xi)}{r^3}\partial_{v_r} F_1  \\ =  \partial_r \psi_1 rg\delta'(v_r)\delta(\ell-\xi) + rg\partial_\theta \psi_1\delta(v_r)\delta'(\ell-\xi), 
\end{gathered}
\end{equation}
with a solution given by an ansatz similar to the 3D case. Indeed, by setting $F_1 = a_1\delta(v_r)\delta(\ell-\xi) + a_2\delta'(v_r)\delta(\ell-\xi)+ a_3\delta(v_r)\delta'(\ell-\xi)$, one can collect the coefficients before the delta distributions and after a Fourier-Laplace transform with $\psi_1 = e^{-i\omega t}\sum_{m=-\infty}^{\infty} \psi_{1,m}(r)e^{im\theta}$, obtain the following system of equations : $i\omega_m a_{1,m} + a_{2,m}' + \frac{im}{r^2}a_{3,m} = 0$ and,
\begin{equation}
\begin{pmatrix}
i\omega_m & \dfrac{2\xi}{r^3} \\
-\xi' & i\omega_m
\end{pmatrix}
\begin{pmatrix}
a_{2,m} \\
a_{3,m}
\end{pmatrix}
=
-rg(r)\begin{pmatrix} 
\psi'_{1,m} \\
i m \psi_{1,m}
\end{pmatrix}.
\end{equation}
Here $\omega_m = \omega - m\Omega$, with $\Omega=\xi/r^2$ being the angular velocity of circular orbits at $r$ and $a_{i,m}$ are the Fourier-Laplace coefficients of the $a_i$ fields. From the 2D Poisson equation $r\psi_{1,m}'' + \psi_{1,m}' - \frac{m^2}{r}\psi_{1,m} = 4\pi G a_{1,m}.$ and after identifying $\kappa^2 = 2\xi \xi' /r^3$ as being the epicyclic frequency, we extract the dispersion relation, which is a single second-order ODE, regular except at the Lindblad radii $\omega_m^2 = \kappa^2$. Near $r = 0$, where $\xi \sim \Omega_0r^2$ and $4\pi G g(0) = 2\Omega_0^2$, solving the matrix gives $a_{2,m} \sim r^\lambda$ and $a_{3,m} \sim r^{\lambda+1}$ with
cross-coefficients $(\bar \omega\lambda - 2m\Omega_0)$ and $(2\Omega_0\lambda - m\bar\omega)$,
where $\bar \omega \equiv\omega_m(0) = \omega-m\Omega_0$; upon forming
$\partial_r a_{2,m} + (im/r^2)a_{3,m} = -i\omega_m a_{1,m}$, the $2m\Omega_0\lambda$ terms
cancel exactly, leaving $a_{1,m} \propto (\lambda^2 - m^2)\,r^{\lambda-1}$.  The indicial equation collapses to
\begin{equation}
\left(1 - \frac{2\Omega_0^2}{\Delta_0}\right)(\lambda^2 - m^2) = 0,
\quad \Delta_0 = 4\Omega_0^2 - \bar\omega^2.
\label{eq:disk_indicial}
\end{equation}
This is the bare 2D Laplacian indicial equation: $\lambda = \pm m$ and is $\omega$-independent. Bar and spiral modes in cold disks are not selected by central regularity~\cite{LinShu1964,Kalnajs1971} but by Lindblad-radius matching. The 3D mode~\eqref{eq:omega0} has no analog here (even with the full 3D Poisson equation).

\textit{Numerical simulations.}---We confirm the mode with nonlinear
Vlasov--Poisson simulations performed with CNVS, a semi-Lagrangian
solver based on the Cheng--Knorr scheme with Strang splitting
(second order in time) and cubic-spline advection, which we validated
against the exact kinetic dispersion relation for nonlinear-wave
sidebands in Ref.~\cite{Tacu2024}. There the code integrated a single
$(x,v)$ Vlasov--Poisson system; here we use that spherical symmetry
conserves the angular momentum $\ell$, so the sphere reduces to
$N_\ell+1$ independent $(r,v_r)$ Vlasov equations at fixed $\ell_k$,
\begin{equation}
\partial_t F_k + v_r\,\partial_r F_k +
\left(\frac{\ell_k^2}{r^3} - \psi'(r, t)\right)
\partial_{v_r} F_k = 0,
\label{eq:vlasov_k}
\end{equation}
with $F_k(r,v_r,t)\equiv F(r,v_r,\ell_k,t)$, coupled only through the
Poisson equation with the spherical density
\begin{equation}
\rho(r) = \frac{2\pi}{r^2}\sum_{k} w_k \ell_k \int dv_r F_k + \int dv_rf_{\rm rad}.
\label{eq:rho_num}
\end{equation}
The $\{\ell_k\}$ are chosen so that the circular radii $r_c(\ell_k)$ given by $\ell_k = \xi(r_c)$, are
equispaced, $w_k$ are trapezoidal weights, and the radial orbits ($\ell=0$) are carried by a
separate pool $f_{\rm rad}(r,v_r)$. 

The equilibrium is a warm distribution $f_0(E,\ell)$ that reduces to the
cold circular-orbit form~\eqref{eq:f0} as $\sigma_E\to0$, built as a
superposition of truncated Boltzmann components
$\propto e^{-(E-E_p)/\sigma_E}\Theta(E-E_p)$, with $E_p$ the energy of
each circular orbit $\ell_k$, supplemented by radial ($\ell=0$)
components spanning the bound energies.  Their non-negative weights are
set by a Tikhonov-regularized ($\lambda=10^{-4}$) least-squares fit of
the density to the Plummer profile, iterated with the Poisson equation
to self-consistency.  The residual is
$\rho_{\rm RMS}\approx1.5\times10^{-4}$ over $r\in[0,2a]$, small enough
that $\kappa(0)=2.0000$ remains the maximum of the epicyclic continuum.

The perturbed force $-\psi_1'$ is advanced self-consistently from
$\rho_1(r,t)=\rho_{\rm num}(r,t)-\rho_{\rm eq}(r)$ by spherical Gauss's
law, with $\rho_{\rm eq}$ the numerical density after a short
pre-relaxation toward the discrete scheme's stationary state rather than
the analytic Plummer profile, so the unperturbed baseline vanishes to
round-off.  The mode is excited by a central radial velocity kick
$\delta v_r(r)=\varepsilon\,v_k\,e^{-r^2/(2r_k^2)}$
($\varepsilon=10^{-2}$, $v_k=0.1$, $r_k=0.3\,a$), driving the orbits
near $r=0$ directly and the rest of the continuum through the
self-consistent potential.  The response is isolated as
$\Delta\psi_1'=\psi_1'(\mathrm{kick})-\psi_1'(\mathrm{ref})$, the
reference run carrying the same residual drift.

\begin{figure}[t]
\centering
\includegraphics[width=\columnwidth]{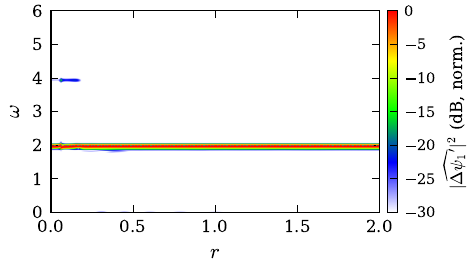}
\caption{%
Power spectrum $|\widehat{\Delta\psi_1'}(r,\omega)|^2$ (dB) of the perturbed radial force on a Plummer equilibrium ($\sigma_E=0.1$, $\varepsilon=10^{-2}$), from a late-window FFT of $\Delta\psi_1'=\psi_1'(\mathrm{kick})-\psi_1'(\mathrm{ref})$, normalized at each radius to its own peak. The bright line at $\omega_0=\kappa(0)=2$ (red dashed: prediction~\eqref{eq:omega0}) lies
at the same frequency for every radius, although the local epicyclic frequency drops to $\kappa(2a)\approx0.4$ at the right edge: the late-time response is a single collective frequency, and not the oscillation at the local epicyclic frequency $\kappa(r)$. The faint feature near $\omega\approx4$ at small $r$ is the $2\omega_0$ nonlinear harmonic.}
\label{fig:spectrogram}
\end{figure}

Production runs use $N_r=1024$, $N_v=512$, $N_\ell=256$ with $n_E=800$
radial pools, $\Delta t=0.05$, integrated to $t=400$, with mass
conserved to better than $10^{-3}$ over the spectral-fit window.  At
radii where the mode is resolved from the continuum ($r\gtrsim0.3\,a$),
the measured peak sits within ${\sim}0.15\%$ of $\kappa(0)$, an offset we
attribute to finite resolution and dispersion. That the peak frequency
is common across these radii, pinned at $\omega_0$ even where
$\kappa(r)$ has fallen well below it (Fig.~\ref{fig:spectrogram}),
identifies it as the discrete kinetic eigenmode~\eqref{eq:omega0}.
\begin{figure}[t]
\centering
\includegraphics[width=\columnwidth]{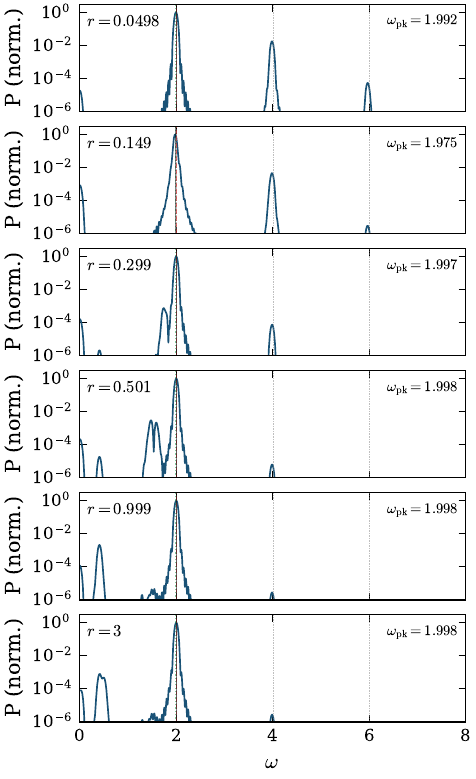}
\caption{%
Late-window power spectra of $\Delta\psi_1'(r,t)$ at six radii
$r\in\{0.05,0.15,0.30,0.50,1.0,3\}$ on a Plummer equilibrium
($\sigma_E=0.1$, $\varepsilon=10^{-2}$), each taken after the initial transient oscillation. At $r\gtrsim0.3$
the peak sits at $\omega_{\rm pk}\approx\omega_0=\kappa(0)=2$ (dashed) to within $0.3\%$, while the local epicyclic frequency falls from $\kappa(0.3)\approx1.8$ to $\kappa(1.5)\approx0.6$. At the innermost radii $\kappa(r)$ approaches $\omega_0$ ($\kappa(0.15)\approx 1.95$), so the mode is no longer spectrally separated from the local epicyclic frequency $\kappa(r)$ and its peak is pulled slightly low ($\omega_{\rm pk}\approx 1.975$ at $r=0.15\,a$). The weaker lines at $2\omega_0,3\omega_0,\dots$ are nonlinear harmonics.}
\label{fig:spectrum}
\end{figure}

\textit{Black hole with stellar cusp.}---With a central black hole,
$\psi_0 = -GM_\bullet/r + \psi_\star(r)$ and
$M_\star' = 4\pi r^2\rho_\star$.  The reduction to $\alpha z=K$
[Eqs.~\eqref{eq:conserved}--\eqref{eq:alpha_def}] used only the
equilibrium Poisson equation; an unperturbed point mass leaves it
intact and enters only through $\Omega$ and $\kappa$.  The self-gravity
numerator becomes
\begin{equation}
\boxed{2\psi_0' + r\psi_0'' = 4\pi Gr\rho_\star,}
\label{eq:identity}
\end{equation}
the Keplerian terms cancelling ($2GM_\bullet/r^2 + r(-2GM_\bullet/r^3)=0$).
With $\kappa^2-\Omega^2 = 4\pi G\rho_\star$ in the potential,
Eq.~\eqref{eq:alpha_ratio} carries over.
Let the circular-orbit population have an inner edge $r_{\rm in}$:
stars occupy $r>r_{\rm in}$, with vacuum inside. The perturbed density vanishes inside and the point mass is fixed, so the perturbed enclosed
mass is zero and $z=0$.  Outside, $\alpha z=K$ gives the perturbed
enclosed mass $r^2 z(r) = K\tfrac{\omega^2-\kappa^2(r)}{\omega^2-\Omega^2(r)}$,
and continuity with the empty interior forces it to vanish at
$r_{\rm in}$.  Since $\omega^2\neq\Omega^2(r_{\rm in})$, this selects
\begin{equation}
\omega_0 = \kappa(r_{\rm in}),\qquad
\kappa^2 = \Omega^2 + 4\pi G\rho_\star.
\label{eq:omega_bh}
\end{equation}
For a centrally concentrated mass $\Omega(r)$ decreases outward, so
$\Omega(r_{\rm in})$ tops the continuum $\{\Omega(r),\,r>r_{\rm in}\}$;
since $\omega_0=\kappa(r_{\rm in})>\Omega(r_{\rm in})$ the mode lies
above it and is undamped, as in the cuspless case.  Its fractional
offset from the circular frequency,
\begin{equation}
\frac{\delta\omega}{\Omega}
= \frac{\kappa(r_{\rm in})}{\Omega(r_{\rm in})}-1
\simeq \frac{2\pi G\rho_\star(r_{\rm in})}{\Omega^2(r_{\rm in})},
\label{eq:shift}
\end{equation}
is positive ($\kappa>\Omega$): the standard retrograde apsidal
precession driven by the enclosed stellar mass, opposite to the
relativistic advance, set by the stellar density at the inner edge; for
a power-law cusp $\rho_\star\propto r^{-\gamma}$ it equals
$\tfrac{1}{2}(3-\gamma)\,M_\star({<}r_{\rm in})/M_\bullet$.  For
Sgr~A$^*$ ($M_\bullet = 4.3\times10^6\,M_\odot$~\cite{gravity2024}), the
S-stars' scale is $r_{\rm in}\sim10^{-2}\,$pc; with a Bahcall--Wolf
slope $\gamma=7/4$~\cite{bahcallwolf1976} and the GRAVITY $1\sigma$
bound $M_\star({<}r_{\rm in})\lesssim1200\,M_\odot$~\cite{gravity2024},
Eq.~\eqref{eq:shift} gives $\delta\omega/\Omega\lesssim2\times10^{-4}$,
$\lesssim4'$ per orbit against the $\approx12'$ Schwarzschild advance of
S2~\cite{gravity2020,gravity2022}---the same retrograde precession the
collaboration already uses to bound the extended
mass~\cite{gravity2024}.  This precession is single-particle and known.
What the kinetic theory adds is the collective mode: a single coherent
oscillation at $\kappa(r_{\rm in})$, the same frequency for stars at
every radius rather than the local $\kappa(r)$.  Were it excited, it
would imprint a common frequency on the astrometric residuals of stars
at different radii, which is a coherence absent from the single-particle precession.  Whether it is excited, and whether that coherence is
observable, are open. Two caveats: the estimate is for near-circular
orbits, and a sharp inner edge is an idealization.

\textit{Discussion.}---We have shown that cold kinetic theory differs from cold fluid theory: in three-dimensional spherical geometry the Vlasov--Poisson system supports a discrete collective mode that the pressureless fluid equations do not see. In the time domain the two descriptions coincide, both
collapsing to independent shells that oscillate at $\Omega(r)$ and phase-mix
away~\eqref{eq:alpha_sol}. They differ only in the spectrum, where the kinetic
system carries one extra solution, at $\omega_0 = \kappa(0) = 2\Omega(0)$.
Sitting above the orbital continuum $\{\Omega(r)\}$, at twice its maximum,
this mode meets no resonance and escapes phase mixing, undamped in the strict
cold limit. Its origin is purely geometric: the angular-momentum factor
$\ell d\ell$ in the spherical velocity measure leaves an integration constant
in the Poisson equation, one that vanishes identically in one dimension and in
razor-thin disks.

The fluid closure misses physics at both extremes of velocity anisotropy. A
radially anisotropic sphere is unstable to perturbations the Jeans equations
cannot see: the radial-orbit instability~\cite{barnes1986}; the
tangentially-supported limit treated here carries, at the other extreme, a
neutral oscillation the fluid cannot see either. The mode is the
self-gravitating counterpart of an isolated oscillation standing above the van
Kampen continuum~\cite{Vandervoort2003}, and it provides in closed form the
discrete solution absent from the spectra of stable warm ergodic
models~\cite{LauBinney2021,Ng2021}.

It may have gone unnoticed because its amplitude peaks at $r = 0$, 
where $N$-body discreteness noise is highest: resolving the ${\sim}2\Omega(0)$
signal calls for a Vlasov integration, as here, or for $N \gtrsim 10^8$
particles with spectral analysis of the central potential. The cold
circular-orbit model is of course an idealization. A finite dispersion
$\sigma$ broadens the line into a resonance of width $\Delta\omega/\omega_0
\sim \sigma/v_\perp$; our simulations show the mode surviving at finite
$\sigma_E$ with its frequency still pinned at $\kappa(0)$
(Fig.~\ref{fig:spectrum}), where it acquires a weak damping whose
physical-versus-numerical origin, together with the warm stability of the
model, we leave open. Applied to Sgr~A$^*$, the model places this mode at
$\kappa(r_{\rm in})$; whether it is excited, and the warm, eccentric
extension the S-stars demand ($e\sim0.3$--$0.9$), are left for future work.
\begin{acknowledgments}
The author thanks Serge Bouquet for discussions on the fluid limit
and Didier B\'{e}nisti for discussions on kinetic theory.
\end{acknowledgments}
 
\bibliographystyle{apsrev4-2}
\bibliography{manuscrit}

@book{binney2008,
  author    = {J. Binney and S. Tremaine},
  title     = {Galactic Dynamics},
  edition   = {2nd},
  publisher = {Princeton University Press},
  address   = {Princeton},
  year      = {2008}
}

@book{peebles1980,
  author    = {P. J. E. Peebles},
  title     = {The Large-Scale Structure of the Universe},
  publisher = {Princeton University Press},
  address   = {Princeton},
  year      = {1980}
}

@article{LinShu1964,
  author  = {C. C. Lin and F. H. Shu},
  title   = {On the spiral structure of disk galaxies},
  journal = {Astrophys. J.},
  volume  = {140},
  pages   = {646},
  year    = {1964}
}

@article{Toomre1964,
  author  = {A. Toomre},
  title   = {On the gravitational stability of a disk of stars},
  journal = {Astrophys. J.},
  volume  = {139},
  pages   = {1217},
  year    = {1964}
}

@article{fillmore1984,
  author  = {J. A. Fillmore and P. Goldreich},
  title   = {Self-similar gravitational collapse in an expanding universe},
  journal = {Astrophys. J.},
  volume  = {281},
  pages   = {1},
  year    = {1984}
}

@article{Ng2005,
  author  = {C. S. Ng and A. Bhattacharjee},
  title   = {Bernstein-Greene-Kruskal modes in a three-dimensional plasma},
  journal = {Phys. Rev. Lett.},
  volume  = {95},
  pages   = {245004},
  year    = {2005}
}

@article{Ng2006,
  author  = {C. S. Ng and A. Bhattacharjee and F. Skiff},
  title   = {Weakly collisional Landau damping and three-dimensional
             Bernstein-Greene-Kruskal modes: New results on old problems},
  journal = {Phys. Plasmas},
  volume  = {13},
  pages   = {055903},
  year    = {2006}
}

@article{Ng2021,
  author  = {C. S. Ng and A. Bhattacharjee},
  title   = {Landau modes are eigenmodes of stellar systems in the limit
             of zero collisions},
  journal = {Astrophys. J.},
  volume  = {923},
  pages   = {271},
  year    = {2021}
}

@misc{Tacu2025,
  author        = {M. Tacu},
  title         = {Why cold {BGK} modes are so cool: Dispersion relations
                   from orbit-constrained distribution functions},
  year          = {2025},
  eprint        = {2504.10382},
  archivePrefix = {arXiv},
  primaryClass  = {physics.plasm-ph}
}

@article{Vandervoort2003,
  author  = {P. O. Vandervoort},
  title   = {On stationary oscillations of galaxies},
  journal = {Mon. Not. R. Astron. Soc.},
  volume  = {339},
  pages   = {537},
  year    = {2003}
}

@article{LauBinney2021,
  author  = {J. Y. Lau and J. Binney},
  title   = {Modes of a stellar system {I}: Ergodic systems},
  journal = {Mon. Not. R. Astron. Soc.},
  volume  = {507},
  pages   = {2241},
  year    = {2021}
}

@article{Kalnajs1971,
  author  = {A. J. Kalnajs},
  title   = {Dynamics of flat galaxies. {I}},
  journal = {Astrophys. J.},
  volume  = {166},
  pages   = {275},
  year    = {1971}
}

@article{barnes1986,
  author  = {J. Barnes and J. Goodman and P. Hut},
  title   = {Dynamical instabilities in spherical stellar systems},
  journal = {Astrophys. J.},
  volume  = {300},
  pages   = {112},
  year    = {1986}
}

@article{bahcallwolf1976,
  author  = {J. N. Bahcall and R. A. Wolf},
  title   = {Star distribution around a massive black hole in a globular
             cluster},
  journal = {Astrophys. J.},
  volume  = {209},
  pages   = {214},
  year    = {1976}
}

@article{gravity2020,
  collaboration = {GRAVITY Collaboration},
  author  = {R. Abuter and others},
  title   = {Detection of the {S}chwarzschild precession in the orbit of
             the star {S2} near the {G}alactic centre massive black hole},
  journal = {Astron. Astrophys.},
  volume  = {636},
  pages   = {L5},
  year    = {2020}
}

@article{gravity2022,
  collaboration = {GRAVITY Collaboration},
  author  = {R. Abuter and others},
  title   = {Mass distribution in the {G}alactic {C}enter based on
             interferometric astrometry of multiple stellar orbits},
  journal = {Astron. Astrophys.},
  volume  = {657},
  pages   = {L12},
  year    = {2022}
}

@article{gravity2024,
  collaboration = {GRAVITY Collaboration},
  author  = {K. {Abd El Dayem} and others},
  title   = {Improving constraints on the extended mass distribution in
             the {G}alactic {C}enter with stellar orbits},
  journal = {Astron. Astrophys.},
  volume  = {692},
  pages   = {A242},
  year    = {2024}
}

@article{Tacu2024,
  title = {Sideband growth rates for differentiable and discontinuous distribution functions},
  author = {Tacu, Mikael and B\'enisti, Didier},
  journal = {Phys. Rev. E},
  volume = {110},
  issue = {4},
  pages = {045205},
  numpages = {17},
  year = {2024},
  month = {Oct},
  publisher = {American Physical Society},
  doi = {10.1103/PhysRevE.110.045205},
  url = {https://link.aps.org/doi/10.1103/PhysRevE.110.045205}
}

@article{Mathur1990,
  author  = {Mathur, S. D.},
  title   = {Existence of oscillation modes in collisionless gravitating systems},
  journal = {Mon. Not. R. Astron. Soc.}, volume = {243}, pages = {529}, year = {1990}
}
 
\end{document}